\documentclass[aps,amsmath,amssymb,prd,superscriptaddress,reprint,nofootinbib,floatfix]{revtex4-1}
\usepackage{indentfirst}
\usepackage{bigstrut}
\usepackage{amssymb}
\usepackage{array}
\usepackage{multirow}
\usepackage{etoolbox}
\usepackage{graphicx}
\usepackage{amsthm}
\usepackage{slashed}
\usepackage{array}
\usepackage{graphicx}
\usepackage{fancyhdr}
\usepackage[table,xcdraw]{xcolor}
\usepackage{float}
\usepackage{subfigure}

\newcommand{\beq}{\begin{equation}}
\newcommand{\eeq}{\end{equation}}
\newcommand{\bea}{\begin{eqnarray}}
\newcommand{\eea}{\end{eqnarray}}
\newcommand{\ba}{\begin{array}}
\newcommand{\ea}{\end{array}}

\def\m1{M_1}
\def\m2{M_2}
\def\m3{M_3}

\def\ch10{\tilde \chi^0_1}

\def\to{\rightarrow}

\newcommand{\lsim}{\mathrel{\mathop{\kern 0pt \rlap
  {\raise.2ex\hbox{$<$}}}
  \lower.9ex\hbox{\kern-.190em $\sim$}}}
\newcommand{\gsim}{\mathrel{\mathop{\kern 0pt \rlap
  {\raise.2ex\hbox{$>$}}}
  \lower.9ex\hbox{\kern-.190em $\sim$}}}

\definecolor{pink}{RGB}{255,105,180}

\begin{document}

\unitlength = 1mm

\title{Probing Flavor Non-Universal Theories Through Higgs Physics \\  at the LHC and Future Colliders}
\author{Wen Han Chiu}
\email{wenhan@uchicago.edu}
\affiliation{Department of Physics, University of Chicago, Chicago, Illinois 60637, U.S.A.}
\author{Zhen Liu}
\email{zliuphys@umd.edu}
\affiliation{Maryland Center for Fundamental Physics, University of Maryland, College Park, Maryland 20742, U.S.A.}
\author{Lian-Tao Wang}
\email{liantaow@uchicago.edu}
\affiliation{Department of Physics, University of Chicago, Chicago, Illinois 60637, U.S.A.}
\affiliation{Enrico Fermi Institute, University of Chicago, Chicago, Illinois 60637, U.S.A}
\affiliation{Kavli Institute for Cosmological Physics, University of Chicago, Chicago, Illinois 60637, U.S.A.}
\date{\today}

\begin{abstract}
We explored the possibility that Higgs coupling to new physics violates flavor universality. In particular, we parameterize such models with dimension-six effective operators which modify the coupling between the first generation quarks, Higgs boson, and $Z$ boson. Through the use of boosted Higgsstrahlung events at both the HL-LHC and potential future hadron colliders, as well as existing ATLAS data for background estimates, we projected constraints on the scale of new physics as a function of the Wilson coefficient. The high energy $Zh$ process will provide unique information about these class of operators, and the sensitivity is competitive with the LEP electroweak precision measurements. We include different scenarios of the overall systematic uncertainties and the PDF uncertainties when presenting the projected sensitivities.  We also discuss the constraints from FCNCs to these flavor-violating models and the complementarity of the exotic Higgs decay to the $Zh$ process.
\end{abstract}

\maketitle

%%%%%%%%%%%%%%%%%%%%%%%%%%%%%%%%%%%%%%%%%%%%%%%%%%%%%%%%%%%%%%%%%%%%%%%%%
%%%%%%%%%%%%%%%%%%%%%%%%%%%%%%%%%%%%%%%%%%%%%%%%%%%%%%%%%%%%%%%%%%%%%%%%%
\section{Introduction}
%%%%%%%%%%%%%%%%%%%%%%%%%%%%%%%%%%%%%%%%%%%%%%%%%%%%%%%%%%%%%%%%%%%%%%%%%
%%%%%%%%%%%%%%%%%%%%%%%%%%%%%%%%%%%%%%%%%%%%%%%%%%%%%%%%%%%%%%%%%%%%%%%%%

The Standard Model (SM) is often perceived as complete with the discovery of the Higgs boson in 2012 \cite{Aad:2012tfa, Chatrchyan:2012xdj}. Below the electroweak (EW) scale, the predictive power of the SM is immense. It provides a mechanism for elementary particles to obtain masses and accurately predicts the rates of particle scattering. However, many puzzles remains to be explained. These include the origin of the electroweak scale and the flavor structure of the SM.
These puzzles indicate the existence of new physics (NP) beyond the Standard Model (BSM).

For any BSM model, once all massive BSM particles above the EW scale have been integrated out, their effects will be encoded in the Wilson coefficients of higher-dimensional operators involving SM particles.
In a flavor-universal theory, there is one dimension-5 operator and 59 dimension-6 operators up to hermitian conjugation \cite{Grzadkowski:2010es}. For most processes accessible at colliders, the leading order correction to the SM is dimension-6.

The Higgs doublet is present in a large number of these operators. Hence, the constraint on the new physics scale, $\Lambda_\text{NP}$, is typically associated with processes involving either the Higgs or the longitudinal modes of the massive gauge bosons. These constraints can be obtained from future Higgs factories, where very clean measurements can be performed. Even in the scenario of cancellations among operators, it is still possible to probe new physics up to $\mathcal{O}(10)$ TeV~\cite{Durieux:2017rsg,Gu:2017ckc,Barklow:2017suo,Fujii:2017vwa,DiVita:2017vrr,Chiu:2017yrx,CEPCStudyGroup:2018ghi, deBlas:2018mhx,An:2018dwb,Abada:2019zxq,deBlas:2019wgy}.

In many of these existing studies and analyses of the Higgs physics at current and future colliders, they tend to focus on universal theories. Especially those involving electroweak precision observables (EWPO). However, in general, most BSM theories have couplings in which the third generation and the first two generations can be considerably different. These include models such as  supersymmetry, composite Higgs, as well as quark flavor models ~\cite{Altmannshofer:2012ar,Low:2015uha,Altmannshofer:2015esa,Evans:2015swa,Bauer:2015fxa,Bauer:2015kzy,Altmannshofer:2016zrn,Bauer:2017cov,Altmannshofer:2017uvs,Gori:2017qwg}.

The constraints from LEP measurements on such new physics scenarios are rather weak. By comparison, the high center-of-mass energy at the Large Hadron Collider (LHC) leads to an enhancement of the new physics effect which scale with a higher power of energy compared to the background~\cite{Farina:2016rws,Azatov:2016sqh,Alioli:2017jdo,Panico:2017frx,
Franceschini:2017xkh,CidVidal:2018eel,Cepeda:2019klc}. This can be further enhanced at future hadron colliders at higher energies, such as the 27 TeV high energy upgrade to the LHC~\cite{Cepeda:2019klc,Abada:2019ono}, and a $pp$ collider in a 100 km tunnel with possible beam center of mass energy at 37.5~TeV~\cite{Mangano:2681366,fcc375} and 100~TeV~\cite{CEPC-SPPCStudyGroup:2015csa,Benedikt:2018csr, Mangano:2681366}. Moreover, the hadronic initial states imply good constraints on light-quark operators by virtue of high statistics from the parton distribution function (PDF). Hence, these hadronic colliders are the best place to search for flavor universality violations. In particular, for operators which modify the couplings between the first generation of quarks associated with the Higgs boson.

In this work, we will focus on probing flavor non-universal theories. We will present results involving the first generation up-type flavor operators, which generally has the best sensitivities at proton-proton colliders. The result can be extended to other operators via the appropriate parton luminosity rescaling and also possibly via the final state selection.

The structure of this paper is as follows: in Sec. \ref{sec:mod}, we will introduce the new physics scenario we are considering in this paper. In Sec. \ref{sec:an}, we will present the projected constraints at both the High Luminosity (HL) LHC and potential future hadron colliders. The possible existing constraints from flavor physics will be discussed in Sec. \ref{sec:fla}. The complementarity of this study with exotic Higgs decay will be discussed in Sec. \ref{sec:exo} and lastly, we will conclude.

\section{Flavor non-universal scenario}
\label{sec:mod}
The flavor non-universal operators in the Warsaw basis associated with the first generation are listed in Table~\ref{tab:operators} \cite{Grzadkowski:2010es}.
\begin{table}[h!]
  \centering
  \begin{tabular}{|c|}
    \hline
    Operators \\ \hline
    $\mathcal{O}_{Hu}= (iH^\dagger\overset\leftrightarrow{D_\mu} H)(\bar{u}_R\gamma^\mu u_R)$ \\
    $\mathcal{O}_{Hd}= (iH^\dagger\overset\leftrightarrow{D_\mu} H)(\bar{d}_R\gamma^\mu d_R)$ \\
    $\mathcal{O}^{(1)}_{HQ}= (iH^\dagger\overset\leftrightarrow{D_\mu} H)(\bar{Q}\gamma^\mu Q) $\\
    $\mathcal{O}^{(3)}_{HQ}= (iH^\dagger\sigma^a\overset\leftrightarrow{D_\mu } H)(\bar{Q}\gamma^\mu\sigma^a Q) $\\
    \hline
  \end{tabular}
  \caption{\label{tab:operators}The set of operators with an energy-enhanced contribution to the $pp\rightarrow Vh,VV$ amplitudes.}\label{Op2}
\end{table}

These operators can be classified using the so-called high energy primaries associated with a given diboson process \cite{Franceschini:2017xkh,Liu:2019vid}. These are the coefficient of the term in the relevant diboson process's signal-to-background ratio with the largest energy scaling behavior. Hence, these are the primary observable in the high-energy limit. So if one wishes to constrain new physics using a diboson process at a hadron collider in a general EFT setup, the leading results in new physics constraints should be associated with one of the operators in Table \ref{Op2}.

The Wilson coefficient of $\mathcal{O}_{Hu}$ is the high-energy primary associated with $f_R\bar{f}_R\rightarrow W_L^+W_L^-$ and $f_R\bar{f}_R\rightarrow Z_L h$. There are existing studies in both of these channels, though only the $WW$ channel has been studied in the flavor non-universal scenario \cite{Grojean:2018dqj, Banerjee:2018bio}.
For the operators $\mathcal{O}_{HQ}^{(1)}$ and $\mathcal{O}_{HQ}^{(3)}$, the contribution to the $WW$ channel is enhanced relative to $Zh$ due to the inclusion of the $t$-channel diagram. As a result, one can expect more stringent constraints on the Wilson coefficients of these operators from the $WW$ process.

To determine the overall reach in the parameter space of non-universal models through $Zh$ production, we focus on the contribution of $\mathcal{O}_{Hu}$. The result of the other operators can be parameterized and derived in a similar manner.

To begin, the effective Lagrangian with dimension-6 operators involving up type quarks is
\begin{equation}
  \mathcal{L}=\mathcal{L}_\text{SM}+\frac{c_{Hu}}{\Lambda^2}(iH^\dagger\overset\leftrightarrow{D_\mu} H)(\bar{u}_{R,i}g_{ij}\gamma^\mu u_{R,j}),
\end{equation}
where $i,j$ are flavor indices. For now we focus on the scenario in which the only nonzero coupling is $g_{uu}=1$. Moving to the mass eigenstate basis, we get
\begin{equation}\label{fla1}
  \mathcal{L}\supset \frac{c_{Hu}}{\Lambda^2}(iH^\dagger\overset\leftrightarrow{D_\mu} H)(\bar{u}'_{R,i}U^\dagger_{R,ij}g_{jk}\gamma^\mu U_{R,kl} u_{R,l}),
\end{equation}
where $U_R$ is the unitary matrix which, alongside $U_L$, diagonalizes the mass matrix. Due to the small charm fraction in the parton distribution functions and the typical smallness of the off-diagonal terms of the rotation matrices in most flavor models, we expect their contributions to be negligible. Hence, we will neglect the contribution from the off-diagonal terms for the $Zh$ process. Moving to the EW broken phase, we have
\begin{equation}\label{fla2}
  i(H^\dagger \overset\leftrightarrow{D_\mu} H)\supset-\frac{g}{2c_w}(v+h)^2Z_\mu.
\end{equation}
This interaction term gives us the relevant Feynman rules for $Zh$:
\begin{equation}\label{frs}
  \begin{aligned}
      \vcenter{\hbox{\includegraphics[height=3 cm]{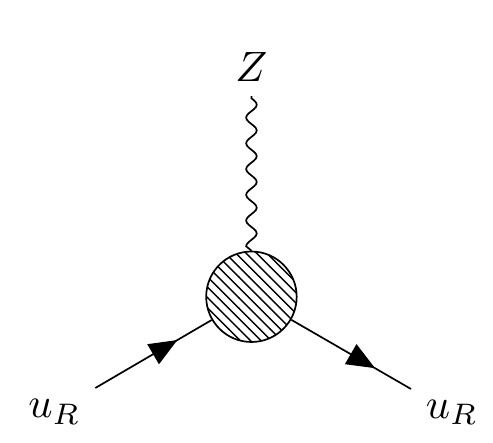}}}&=-i\frac{gv^2}{2c_w}\frac{c_{Hu}}{\Lambda^2}\\
      \vcenter{\hbox{\includegraphics[height=3 cm]{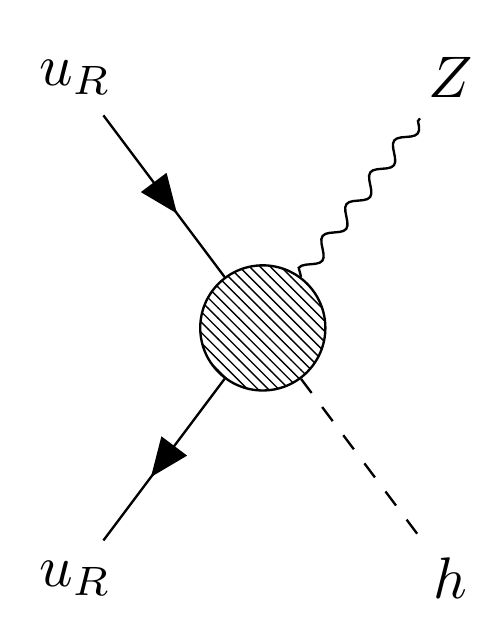}}}&=-i\frac{gv}{c_w}\frac{c_{Hu}}{\Lambda^2}
  \end{aligned}
\end{equation}
where $c_w$ and $s_w$ (in later text) denote $\cos\theta_w$ and $\sin\theta_w$ of the Weinberg angle $\theta_w$ with $s_w^2\simeq0.23$, and $g$ is the $SU(2)$ gauge coupling.

\section{Detailed Analysis}
\label{sec:an}
To obtain projections on the sensitivities, one million $pp\rightarrow Zh$ events were generated in \textsc{MG5\_aMC} with the operator implemented using a \textsc{UFO} file generated in \textsc{FeynRules} \cite{Alwall:2014hca, Degrande:2011ua, Alloul:2013bka}. The Wilson coefficient normalized with a NP scale of 1 TeV, $c_{Hu}/\Lambda^2_\text{TeV}$, was varied from -1 to 1 in increments of 0.1. The data were then scaled to match the number of expected events for a given integrated luminosity. Next, the signal was split into bins of 150 GeV, matching roughly the energy resolution of the $Zh$ system invariant mass over a large range. The number of signal events as a function of the Wilson coefficient was obtained by interpolation.

The SM background under 3 TeV was estimated using the 2017 ATLAS search on heavy resonances to $Zh$ final state \cite{Aaboud:2017cxo}. Above the 3 TeV threshold, the background was modeled by fitting the tail of the data to an exponential function, equivalent to a fixed selection efficiency for high invariant mass regions of the background.

Our signal $Z$ and $h$ with subsequent decays into dileptons and $b\bar{b}$ were multiplied by the corresponding decay branching fractions respectively, to match the final state of the ATLAS search. A $p_T>300~\text{GeV}$ cut and a $|\eta|<2.5$ cut were applied to the Higgs. A universal cut efficiency was then imposed on the signal events to match the number of Standard Model events computed in the ATLAS search.

A binned likelihood test was performed by defining the significance, $Z$, of each bin as a function of the Wilson coefficients, e.g., $c_{Hu}/\Lambda^2_\text{TeV}$ for signal and background numbers of events of $s$ and $b$, as
\begin{equation}\label{sig}
\resizebox{\columnwidth}{!}{$
  Z_i=\left[2\left((s+b)\ln\left[\frac{(s+b)(b+\delta_b^2)}{b^2+(s+b)\delta_b^2}\right]-\frac{b^2}{\delta_b^2}\ln\left[1+\frac{\delta_b^2 s}{b(b+\delta_b^2)}\right]\right)\right]^{1/2},$}
\end{equation}
where $\delta_b$ is the uncertainty~\cite{Cowan:2010js}. The 2$\sigma$ constraint up to a given center-of-mass energy, $\sqrt{\hat{s}}$, was computed by adding the significance of bins with $m_{Zh}<\sqrt{\hat{s}}$ in quadrature and solving for $\sum_i Z_i^2(c_{Hu}/\Lambda^2_\text{TeV})=4$.

When calculating the sensitivities for future hadron collider, signal events were obtained in the same manner as the HL-LHC calculation. For the background, a differential rescaling was performed by computing the ratios of the parton luminosity at each mass bin using \textsc{ManeParse~2.0}~\cite{Clark:2016jgm} and the \textsc{NNPDF23\_nlo} PDF set~\cite{Ball:2012cx} yielding a background estimate for $\sqrt{\hat{s}}<14~\text{TeV}$. As the effective theory is only well-defined for energy scales below the cut-off, our constraints are physically meaningful if $\hat{s}<\Lambda^2$. Hence, only bins with $\hat{s}<\Lambda^2$ will be used in calculating the sensitivities.

\begin{figure}[h!]
  \centering
  \includegraphics[width=\columnwidth]{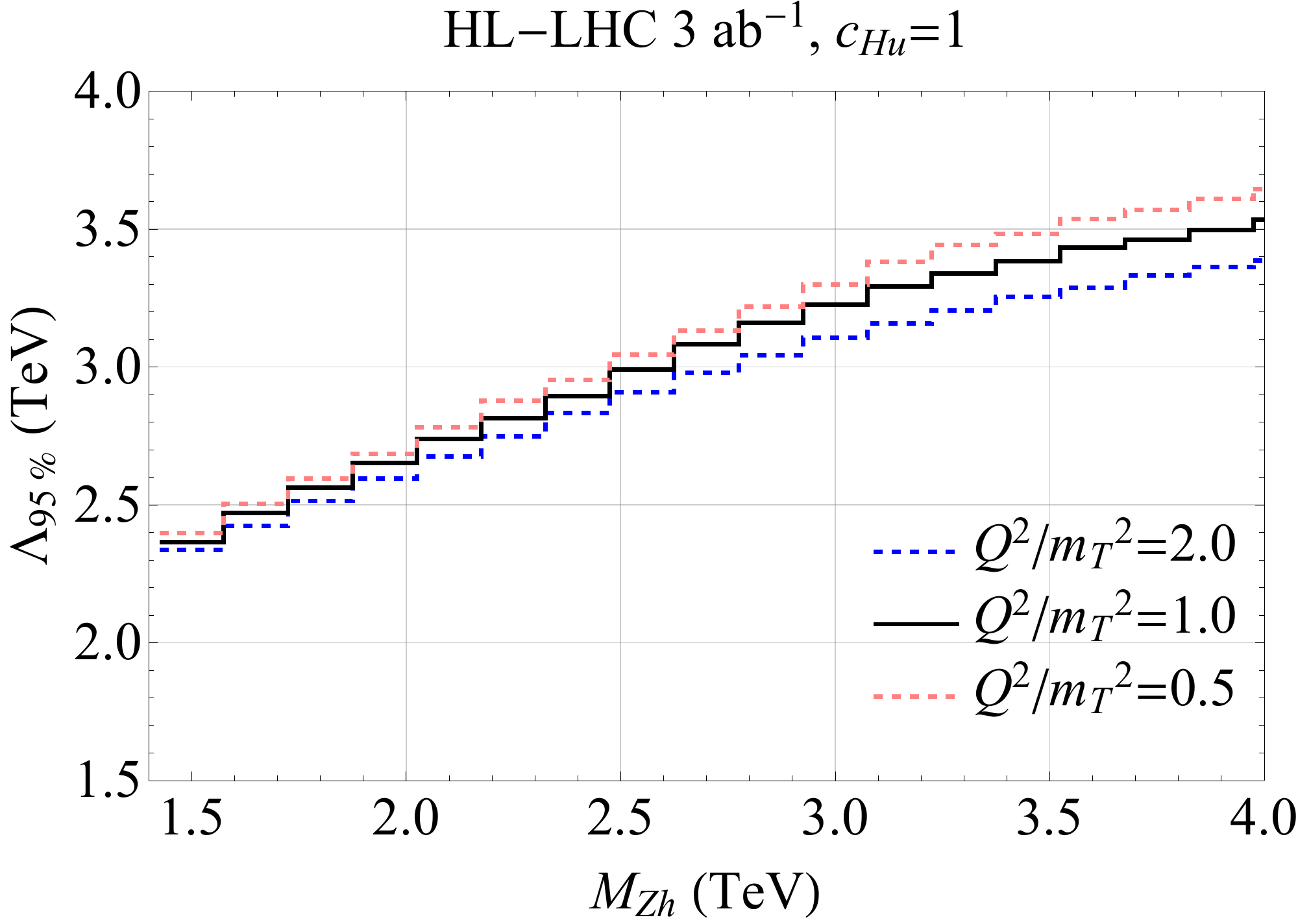}
  \caption{The effects of changing the scale factor on the constraints from HL-LHC with a reference value of $c_{Hu}=1$.
  }\label{fig:bands}
\end{figure}

For the uncertainty used in our analysis, a 5\% universal systematic and statistical uncertainty was assumed.
However, it should be noted that for bins with larger invariant masses, the theoretical uncertainty from the choice of factorization scale increases. This increase of uncertainty can be estimated by performing the analysis with the factorization and normalization scale set to be 0.5, 1, and 2 times $m_T^2$, where, $m_T$ is the transverse mass of the system. We show the sensitivity to the new physics scale with these different choices of the scales in Fig.~\ref{fig:bands}. The scale dependence of our sensitivity as discussed earlier, grows with center of mass energy, up to roughly 3\% with the $Zh$ center of mass energy at 4 TeV.

To take the scale dependence into account, we assumed the per bin PDF uncertainty is Gaussian and defined it as:
\begin{equation}\label{PDF}
\resizebox{\columnwidth}{!}{$
  \delta _{\text{PDF},i}=\frac{1}{2}\Big(\Big|n_i^{\frac{Q^2}{m_T^2}=1}-n_i^{\frac{Q^2}{m_T^2}=0.5}\Big|+\Big|n_i^{\frac{Q^2}{m_T^2}=1}-n_i^{\frac{Q^2}{m_T^2}=2}\Big|\Big),$}
\end{equation}
where $n_i$ is the number of events in the $i$th bin after imposing the appropriate cuts. This was added to our systematic uncertainty linearly.

\begin{figure}[h!]
  \centering
  \includegraphics[width = \columnwidth]{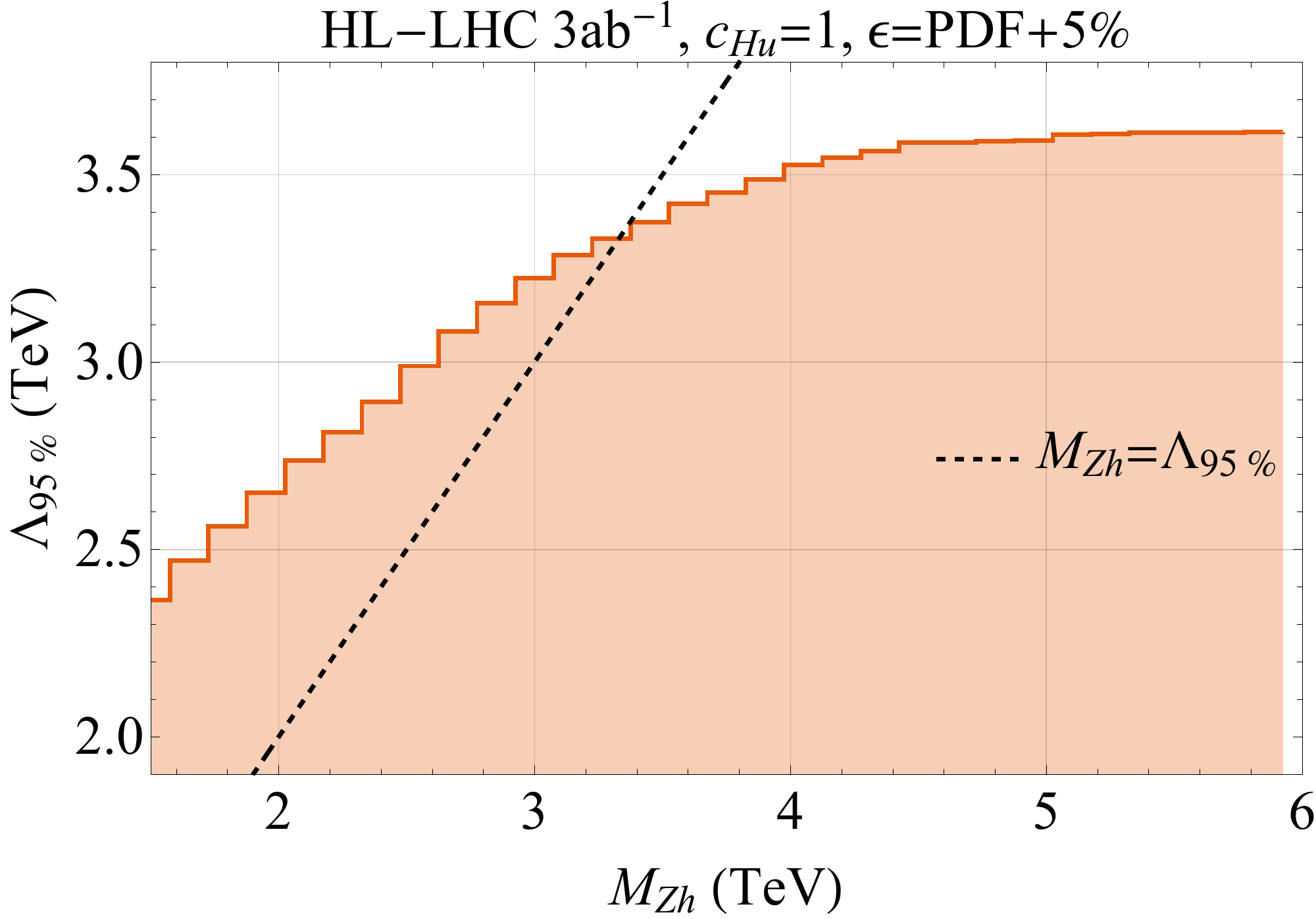}
  \includegraphics[width = \columnwidth]{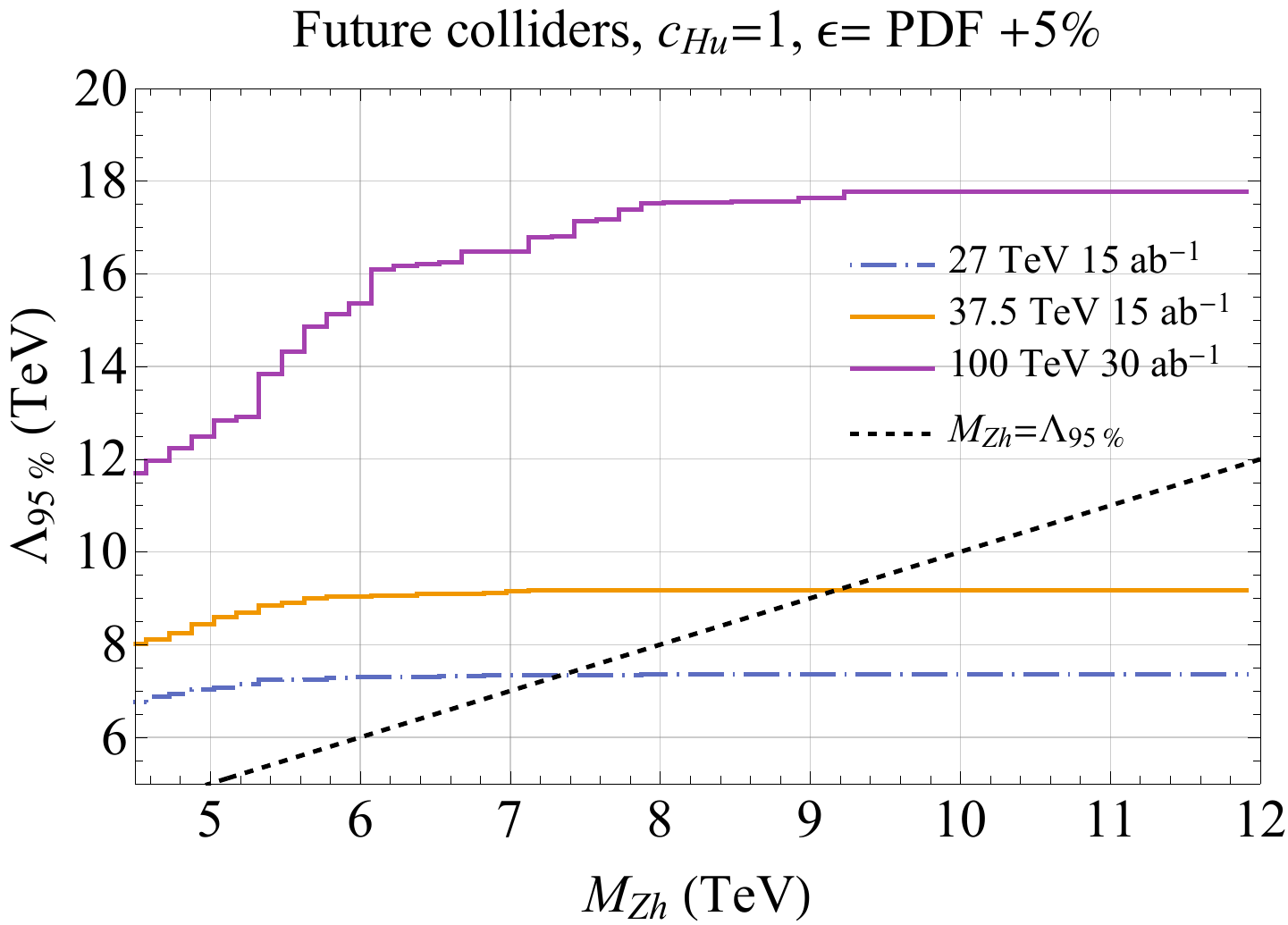}
  \caption{The constraints on the scale of new physics including PDF uncertainties probed by HL-LHC (above) and potential future hadron colliders (below) using the benchmark point of $c_{Hu}=1$.}\label{result}
\end{figure}

The 95\% C.L. sensitivity for the benchmark Wilson coefficient, $c_{Hu}=1$, including the PDF uncertainties, are given in Fig. \ref{result}.
Constraints on new physics scales up to about 3.3 $\text{TeV}$ can be obtained for HL-LHC, 7.3 $\text{TeV}$ for HE-LHC, 9.2 TeV for a 37.5 TeV FCC-hh, and 17.8 TeV for a 100 TeV FCC-hh.

\begin{figure}[h]
  \centering
  \includegraphics[width = \columnwidth]{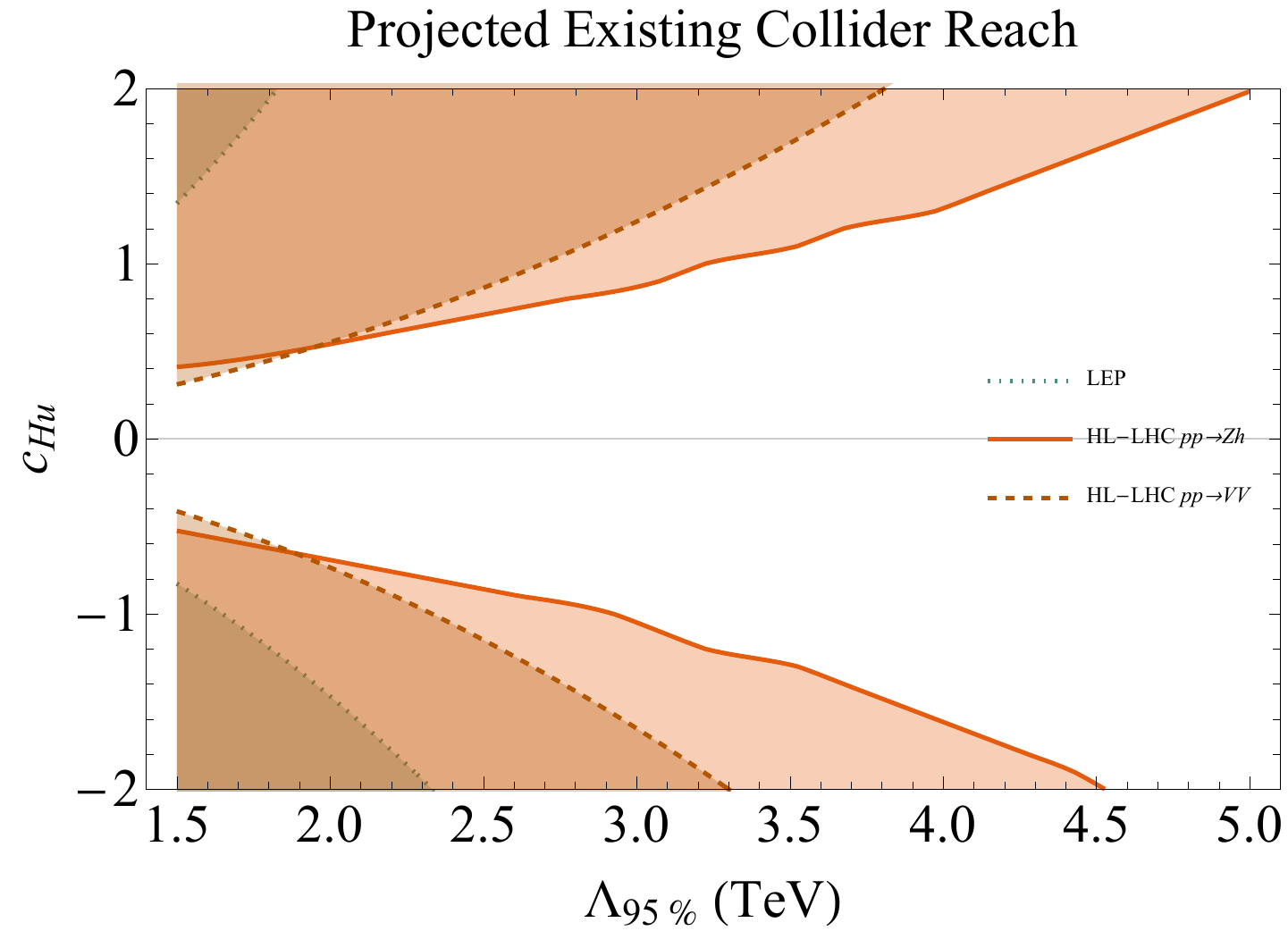}
  \includegraphics[width = \columnwidth]{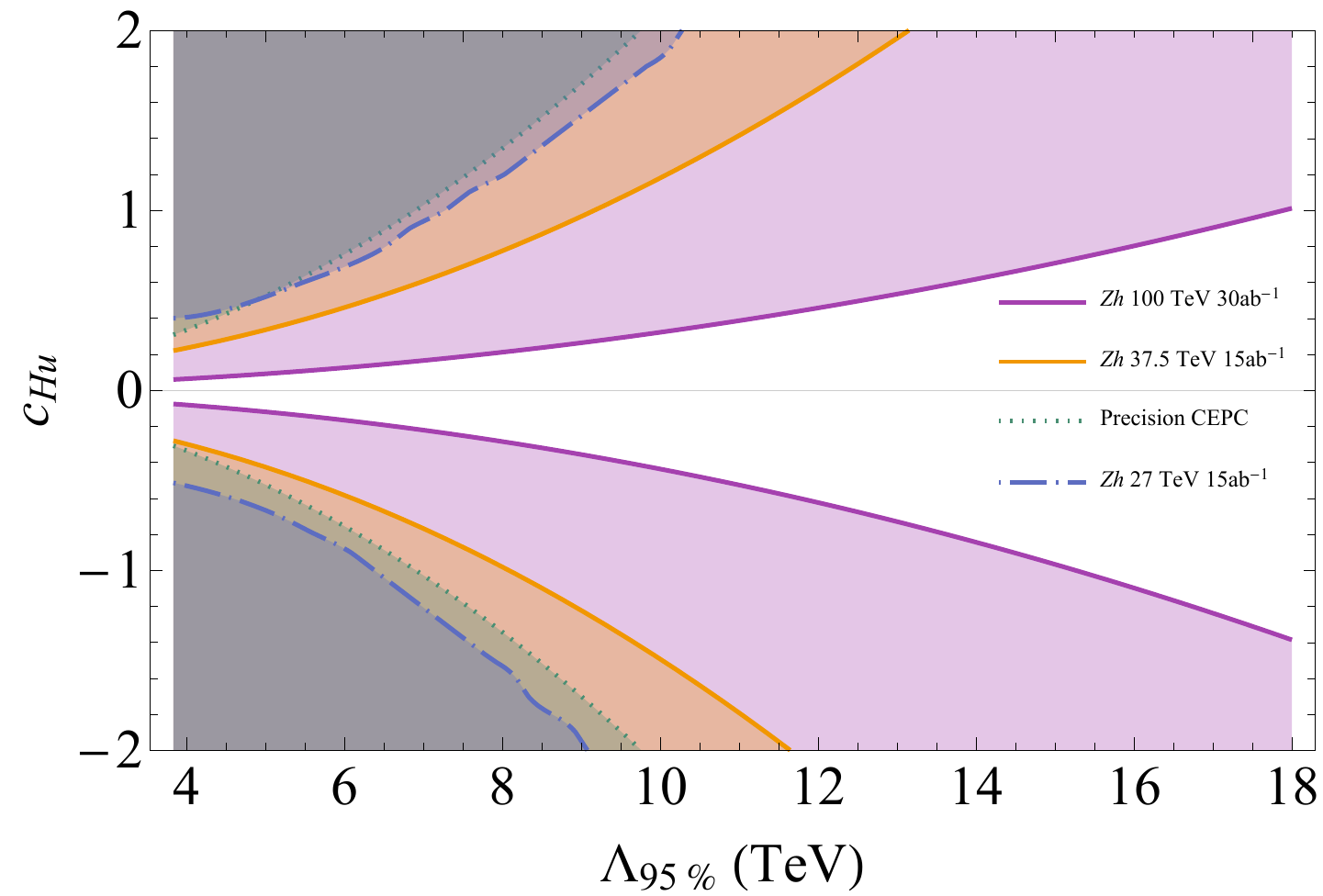}
  \caption{The constraints on the scale of new physics for models with different values of $c_{Hu}$ which can be probed by Higgsstrahlung at HL-LHC (above) and potential future hadron colliders (below) using only bins satisfying $\hat{s}<\Lambda^2$. For comparison, the existing constraints from LEP and the reach from diboson at HL-LHC was also included for the HL-LHC plot while a next generation lepton collider reach was included for the future collider plot.
%A portion of the two FCC-hh scenarios is extrapolated from the simulated data where the background behavior is better modeled.
}\label{final}
\end{figure}

For regions of parameter space beyond our benchmark point, we redid the calculations with different values of $c_{Hu}$. The regions of parameter space that can be probed are given in Fig. \ref{final}. For comparison, the constraints from a corresponding lepton collider were also included in the plot. The LEP constraints were obtained by looking at the shift in $g_R^{Z,u}$ induced by our operator and fitting to the number provided in Ref.~\cite{Efrati:2015eaa}. The CEPC projections were obtained from Ref.~\cite{deBlas:2019wgy}, assuming flavor universality. This assumption will result in a more optimistic estimate as the EWPO will also receive contributions from the other generations.
In addition, we also include the reach from $pp\rightarrow WW$ for HL-LHC by translating the constraint on $g^{Z,u}_R$ into a constraint on $c_{Hu}/\Lambda^2$ \cite{Grojean:2018dqj}. From the figure, we can see that $Zh$ production is indeed competitive to other direct and indirect probes over a large range of parameter space.
%...

To ensure that neglecting dimension-8 operators is well justified, its contribution must be small relative to the dimension-6 operators. First, we can compute the change in the invariant matrix element in powers of $\hat{s}$.
\begin{equation}\label{mat}
\resizebox{\columnwidth}{!}{
  $\frac{\Delta|\overline{\mathcal{M}}|^2}{|\overline{\mathcal{M}}_\text{SM}|^2}=\frac{s_w^4c_w^2}{e^2(32s_w^4-24s_w^2+9)}\left(144\frac{c_w^2}{e^2}\frac{c_{Hu}^2}{\Lambda^4}\hat{s}^2+96\frac{c_{Hu}}{\Lambda^2}\hat{s}\right)$},
\end{equation}
where $e$ is the electric coupling constant around $0.3$.

Noting that the coefficient of the quadratic piece is an order of magnitude larger than the linear piece, the contribution from $|\mathcal{O}_{d=6}|^2$ will dominate once $c_{Hu}\hat{s}/\Lambda^2\gtrsim\mathcal{O}(0.1)$. Given the same suppression of $\Lambda^4$, the contribution from dimension-8 operators should be estimated as well.

As dimension-8 operators do not generate any new vertices which contribute to $Zh$ production at tree-level, they contribute by modifying the vertex factors in Eq. (\ref{frs}). So one can estimate the leading contribution by taking the linear piece in Eq. (\ref{mat}) and replacing
\begin{equation}
  c_{Hu}\rightarrow c_{Hu}+\sum_{i,j} a_{i}c_{i,j}\frac{p_i\cdot p_j}{\Lambda^2},
\end{equation}
where the $i,j$ indices denote the different legs in the Feynman diagram and $a_{i}$ is some $\mathcal{O}(1)$ number.

So, in models where the Wilson coefficients of the dimension-8 operators are less than or comparable to the dimension-6 operators, the leading contribution from dimension-8 are estimated to be smaller than dimension-6 and dimension-6 squared.
In cases where dimension-8 operators Wilson coefficients being larger than dimension-6, one should view our constraints as those on a given linear combination of the Wilson coefficient of dimension-6 and dimension-8 that can be absorbed into the dimension-6 operators. For instance, dimension-8 operators derived with additional $H^\dagger H$ insertions to the dimension-6 operators can be captured by redefining the dimension-6 operators' coefficients concerning the $Zh$ process considered in this work. The estimation of the sensitivity to new physics scale $\Lambda$ for Wilson coefficient of order unity remains the same.

\section{Flavor Physics Constraints}
\label{sec:fla}
The type of flavor models that we are looking at may have non-trivial constraints from flavor changing neutral currents (FCNC). This is due to the presence of flavor-mixing terms in the Lagrangian in the mass eigenbasis. The dominant constraint on up-type flavor mixing is through charm-number violating processes, in particular from $D_0-\bar{D}_0$ mixing \cite{Gedalia:2009kh}. In order to have a rough estimate of what region of parameter space has been ruled out by existing measurements, we computed the leading order contribution from our operator.

From Eqs. (\ref{fla1}) and (\ref{fla2}), the operator which directly contributes to FCNC via $D_0-\bar{D}_0$ mixing is
\begin{equation*}
  \mathcal{L}^{\Delta C=1}=-\frac{c_{Hu}M_Zv}{\Lambda^2}Z_\mu \bar{u}_R\gamma^\mu c_R (U^\dagger_{R,uu}U_{R,uc}).
\end{equation*}
Integrating out the $Z$ propagator gives the effective operator:
\begin{equation*}
  \mathcal{L}^{\Delta C=2}_\text{eff}=3\left(\frac{c_{Hu}v}{\Lambda^2}(U^\dagger_{R,uu}U_{R,uc})\right)^2\bar{u}_R\gamma^\mu c_R\bar{u}_R\gamma_\mu c_R.
\end{equation*}

\begin{figure}[ht]
\centering
\includegraphics[width=\columnwidth]{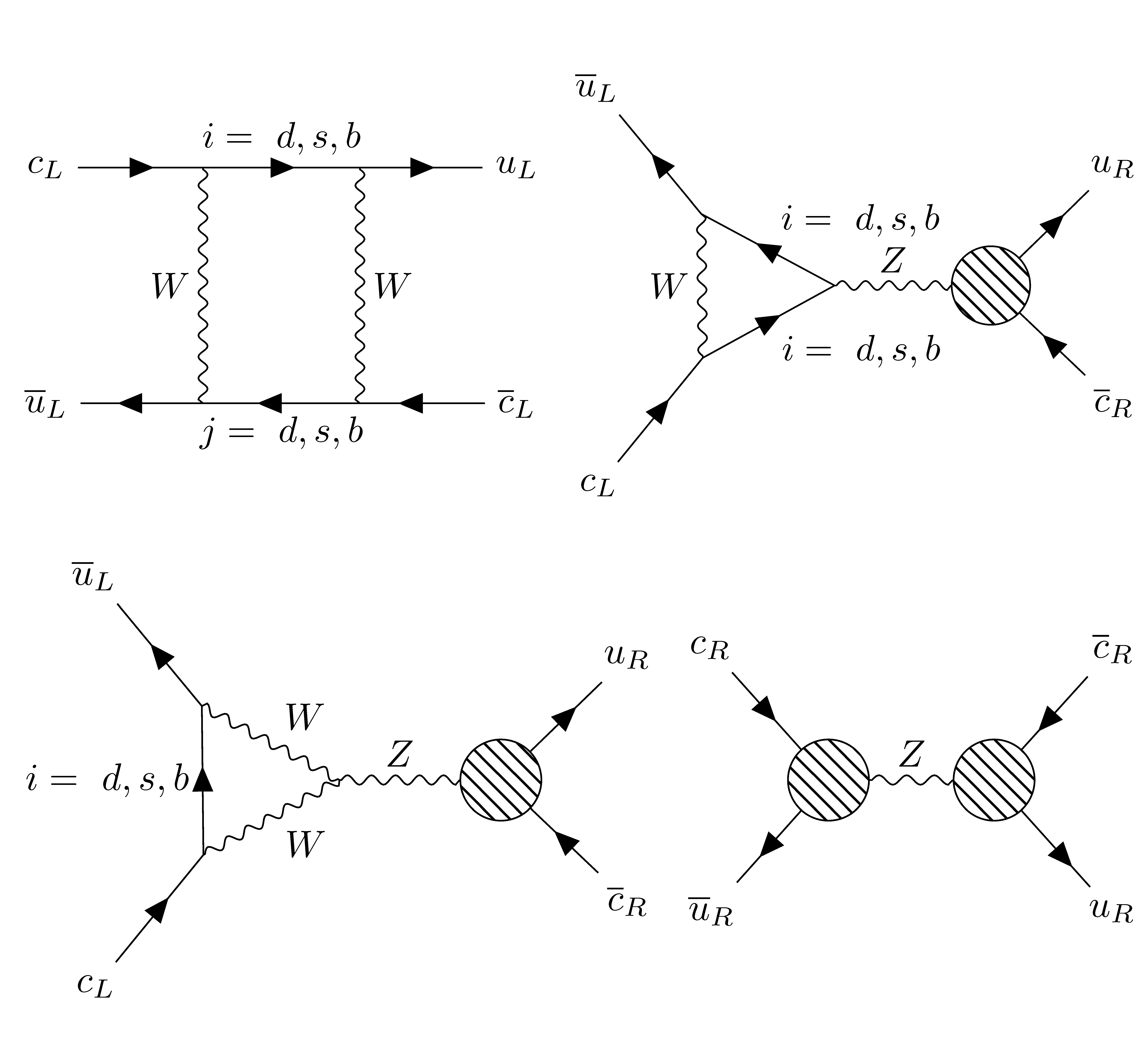}
\caption{A subset of diagrams which contribute to $D_0-\overline{D_0}$ mixing at leading order to illustrate the parametric dependence for each $\Delta C=2$ effective operators.
}\label{fd2}
\end{figure}

From Fig. \ref{fd2}, it can be seen that at dimension-6, SM process contributes to the operator $\bar{u}_L\gamma^\mu c_L\bar{u}_L\gamma_\mu c_L$ while the SM-EFT cross terms contributes to operators of the form $\bar{u}_L c_L \bar{u}_R c_R$. Thus, the leading contribution to the operator $\bar{u}_R\gamma^\mu c_R\bar{u}_R\gamma_\mu c_R$ is the EFT only term. The Wilson coefficient of this operator has been constrained in \cite{Gedalia:2009kh} using a global fit with all possible low-energy dimension-6 operators. As such, this should be viewed as a conservative estimate of the current constraint. Assuming that only our operator contributes to the observables used to derive these constraints, we obtain
\begin{equation}\label{flavcon}
  3\left|\frac{c_{Hu}v}{\Lambda^2}(U^\dagger_{R,uu}U_{R,uc})\right|^2\lesssim 5.7\times10^{-7}\left(\frac{1}{1 \text{ TeV}}\right)^2.
\end{equation}
This gives us a constraint on the Wilson coefficient of the operator that depends on the flavor model of interest. As a benchmark model (benchmark theory 1), suppose that $|U^\dagger_{R,uu}U_{R,uc}|=|V_{ud}V_{cd}|$, we get
\beq
\Lambda/\sqrt{c_{Hu}}\gsim 11~{\rm TeV}.
\eeq

In addition, one could consider constraints from the operator $\bar{u}_L c_L \bar{u}_R c_R$. The dominant contribution comes from the bottom quark, so Wilson coefficient is on the order of
\begin{equation}
\begin{aligned}
  &\sim\frac{v^2}{M_Z^2}\frac{M_b^2}{M_W^2}\frac{1}{16\pi^2}\frac{c_{Hu}}{\Lambda^2}|V_{ub}||V_{cb}|(U^\dagger_{R,uu}U_{R,uc})\\
  &\lesssim 1.6\times 10^{-7}\left(\frac{1}{1\text{ TeV}}\right)^2.
\end{aligned}
\end{equation}
The constraints on our Wilson coefficient from this operator in benchmark theory 1 is $\Lambda/\sqrt{c_{Hu}}\gsim 0.16~{\rm TeV}$, clearly weaker than the previous one.

\begin{figure}
  \centering
  \includegraphics[width=\columnwidth]{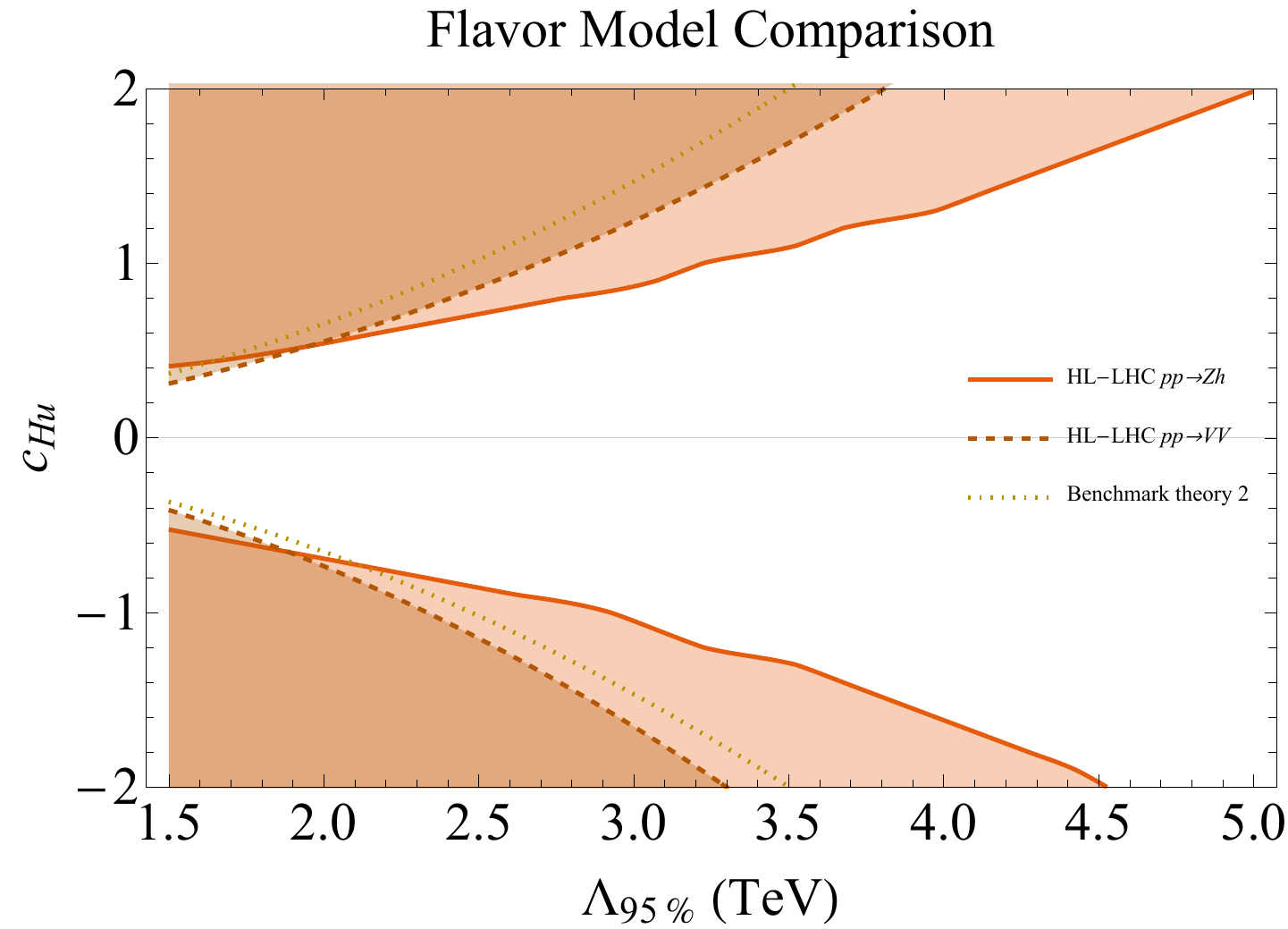}
  \includegraphics[width=\columnwidth]{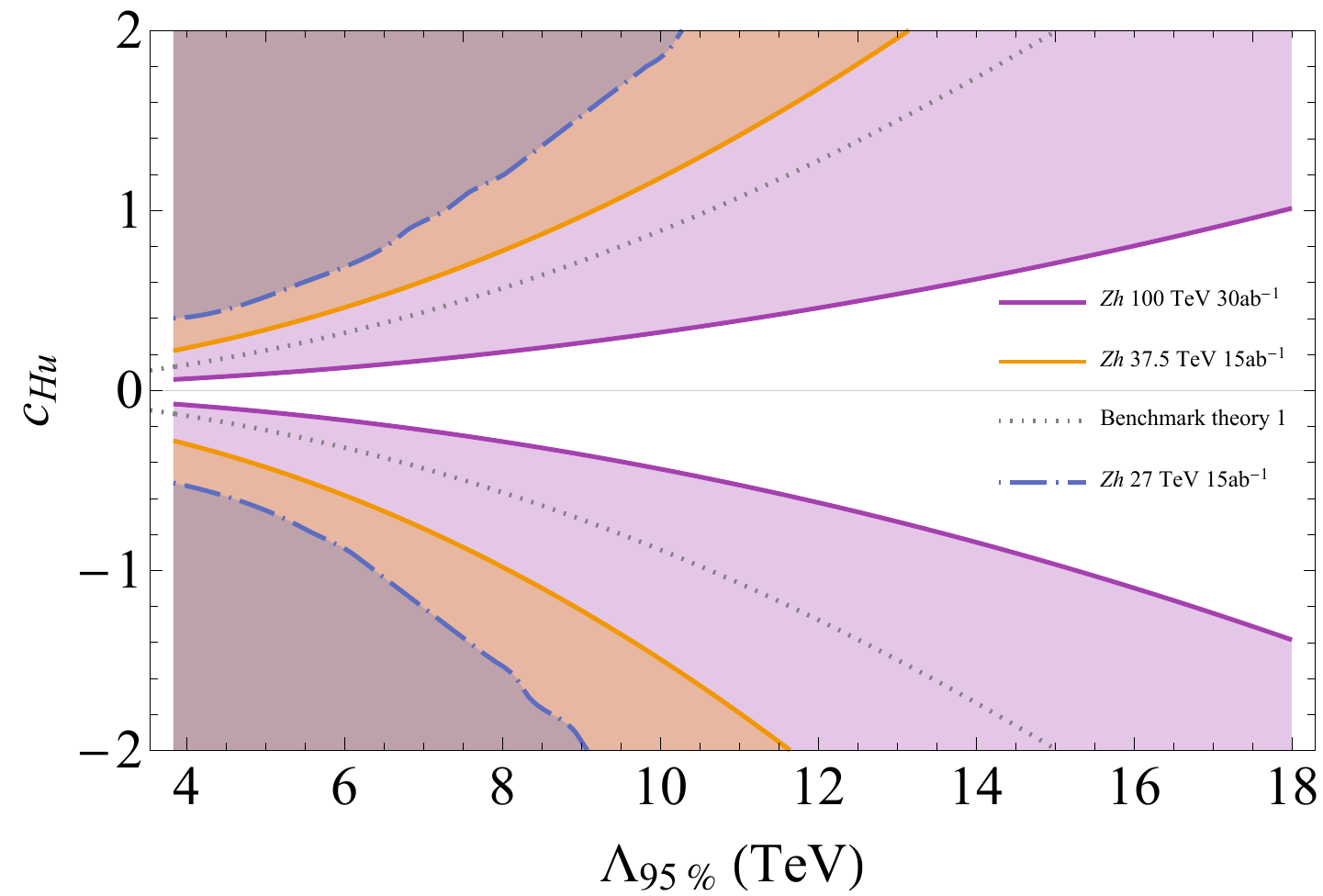}
  \caption{The flavor constraints (unfilled, dotted curves) plotted on top of the HL-LHC (top) and potential future hadron collider (below) constraints. Benchmark theory 1 refers to the fully flavor non-universal theory with the choice of $|U^\dagger_{R,uu}U_{R,uc}|=|V_{ud}V_{cd}|$. Benchmark theory 2 refers to the theory which is universal across the first 2 generations with a similar choice for the right-handed rotations.
  }\label{flavcons}
\end{figure}

These flavor constraints appear to be very strong in a generic flavor violating theory. However, we can consider models where the operator under considerations applies to the first two generations universally, maintaining a $U(2)^2$-flavor symmetry in the quark sector (benchmark theory 2). These types of models can be motivated due to the large mass gap between the second and third generation \cite{Altmannshofer:2017uvs, Gori:2017qwg}. In this scenario, $g_{ij}=\text{diag}(1,1,0)$. The LHC constraints are not expected to change by much due to the limited charm fraction in the large $x$ region. In the mass eigenstates, we now have

\begin{equation*}
\resizebox{\columnwidth}{!}{$
  \mathcal{L}^{\Delta C=1}\rightarrow-\frac{c_{Hu}M_Zv}{\Lambda^2}Z_\mu \bar{u}_R\gamma^\mu c_R \left(U^\dagger_{R,uu}U_{R,uc}+U^\dagger_{R,uc}U_{R,cc}\right).$}
\end{equation*}

 As $U_R$ is unitary, the term in the parenthesis is equal to $-U^\dagger_{R,ut}U_{R,tc}$. Due to the smallness of the corresponding CKM elements, this quantity is naturally small in most flavor models. This relaxes the constraints given by Eq.~(\ref{flavcon}) by a factor of about $\mathcal{O}(10^{-3})$ for a similar benchmark point and hence relax the constraints on the Wilson coefficient to be
\beq
\Lambda/\sqrt{c_{Hu}}\gsim 0.4~{\rm TeV}.
\eeq

The $U(2)$-flavored quark sectors also modifies $\Gamma(Z\rightarrow c\bar{c})$; which has been measured to about 1.6\% accuracy~\cite{Tanabashi:2018oca}. In the small charm mass limit, the fractional change in the $Z\rightarrow c\bar{c}$ width is given by:

\begin{equation}\label{zwidth}
  \frac{\Delta\Gamma(Z\rightarrow c\bar{c})}{\Gamma(Z\rightarrow c\bar{c})}\approx\frac{2g_R\Delta g_R}{g_L^2+g_R^2}\approx -0.0615\frac{c_{Hu}}{\Lambda^2_\text{TeV}}
\end{equation}

For comparison, constraints for the benchmark flavor models were plotted on top of the collider constraints.
From Fig. \ref{flavcons}, the reach from $WW$ production is comparable to the region ruled our by FCNCs for the partial universal theory (benchmark theory 2). Higgsstrahlung at a 37.5 TeV $pp$ collider with 15 ab$^{-1}$ of integrated luminosity is comparable to the region ruled out by existing FCNC measurements for the fully flavor non-universal theory (benchmark theory 1).

\section{Complementarity to Higgs Exotic Decay}
\label{sec:exo}
\begin{figure}[t]
  \centering
  \includegraphics[width=\columnwidth]{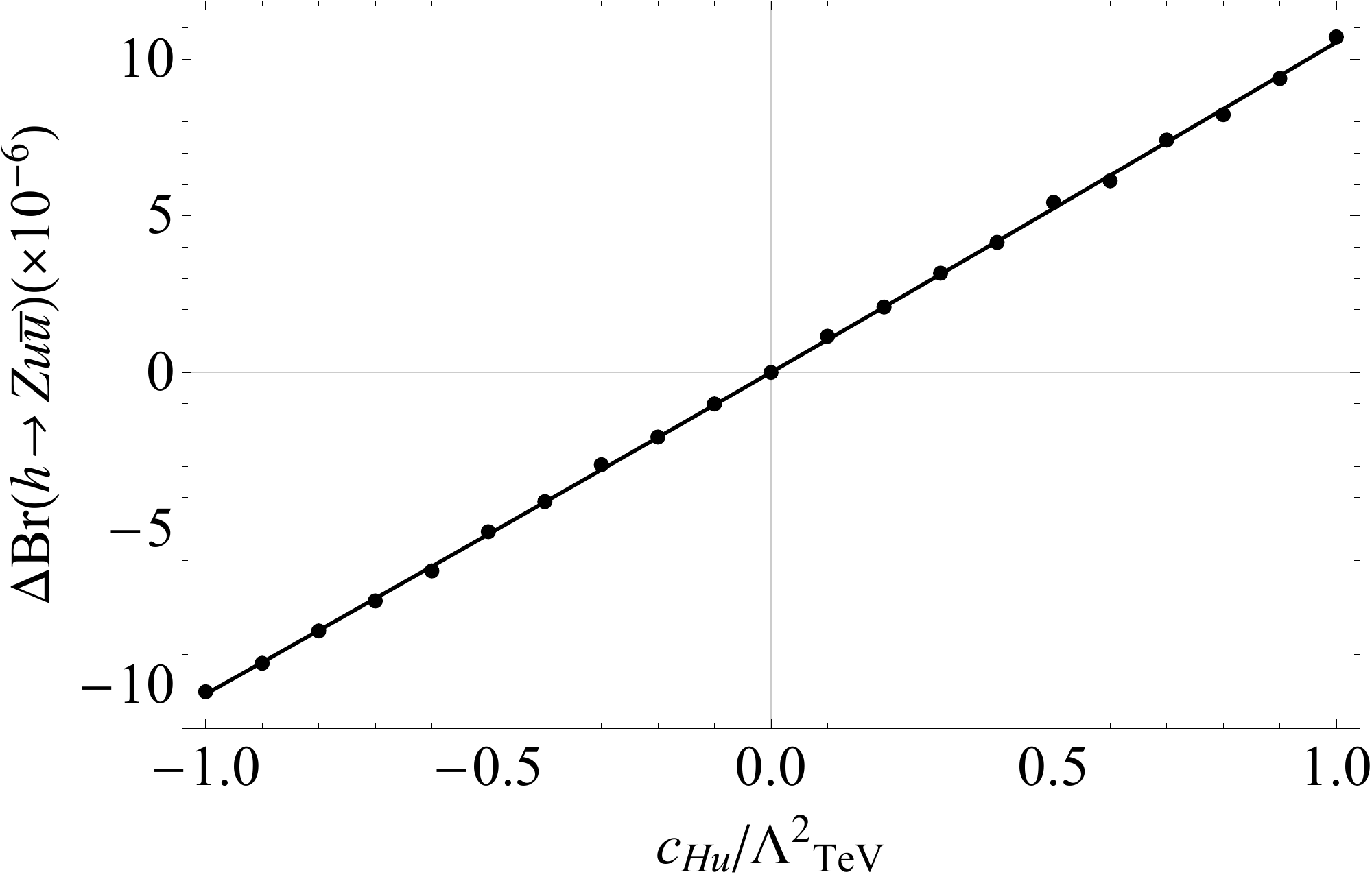}
  \caption{The modifications to the Higgs decays into an on-shell Z boson and quark anti-quark pairs from the operator under consideration.}\label{BR}
\end{figure}

\begin{figure}[ht]
  \centering
  \includegraphics[width=\columnwidth]{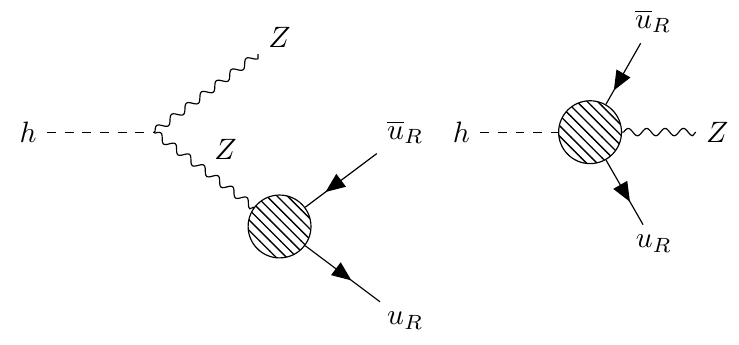}
  \caption{The additional diagrams contributing to $h\rightarrow Zu\bar{u}$.}\label{fd5}
\end{figure}

The Higgs physics exotic decays~\cite{Liu:2016zki} are also modified by these operators.
The operator $\mathcal{O}_{Hu}$ contributes directly to $h\rightarrow Zu\bar{u}$ decay through the addition of the two diagrams in Fig. \ref{fd5}. The first by shifting $g_R$ and the second by generating a contact term.

The shift on the branching ratio was computed using the same model file with \textsc{MG5\_aMC}, shown in Fig.~\ref{BR} as a function of the Wilson coefficient over operator scale squared. We can see the interference term dominance and generically the shift of the order $10^{-5}$-$10^{-6}$. The future lepton collider Higgs factories will produce around one million Higgs bosons in a clean environment. In principle, the modification can be measured as an exclusive mode, especially with charm-quark flavor tagging. Furthermore, in contrast to the $H\to ZZ^*$, this channel would exhibit different kinematic features.

Assuming an upper limit in this channel of $10^{-5}$ and $3\times 10^{-6}$, we can probe $c_{Hu}/\Lambda^2_\text{TeV}$ up to order unity and 0.3, respectively. With a dedicated search, this may further improve. Although not competitive to the high energy probes, this channel does provide a complementary probe to the same physics and will help reveal the nature of the underlying physics.

\section{Conclusion and outlook}
\label{sec:conclusion}
By parameterizing the effects of new physics with non-renormalizable operators, we have studied the potential reach of the HL-LHC and future colliders which modify the $Z$, $h$, and quark couplings in flavor non-universal models. Using a binned-likelihood test, we determined that $Zh$ production is the optimal diboson process to yield constraints on the dimension-six operator, $\mathcal{O}_{Hu}$ and $\mathcal{O}_{Hd}$, beating the constraints from LEP and the $WW$ production at the HL-LHC. With a detailed analysis, we computed the projected sensitivity of the $Zh$ process on these operators at the HL-LHC, HE-LHC, and FCC-hh. In comparison with the flavor constraints and future lepton collider projection, our results show that the $Zh$ process yields competitive sensitivities .

Depending on the choice of right-handed rotations, a portion of the parameter space for flavor non-universal models not excluded by existing FCNC measurements can be tested by the Higgsstrahlung process.  For instance, Higgsstrahlung can exclude a significant portion of the first 2 generation partial universal theories compared with what is currently excluded by measurements. This study also shows the exotic Higgs decay searches at future Higgs factories are complementary to the high energy $Zh$ process. Should future measurements establish any deviations in quark couplings, our proposed measurements will help reveal the flavor nature of the underlying new physics.

\begin{acknowledgements}
The authors would like to thank Emmanuel Stamou and Da Liu for helpful discussions.
ZL and LTW would like to thank Aspen Center for Physics (Grant \#PHY-1607611), KITP, IHEP, KAIST and MIAPP physics programs for support and providing the environment for collaboration during various stages of
this work. ZL is supported in part by the
NSF under Grant No. PHY1620074 and by the Maryland Center for Fundamental Physics. LTW is supported by the DOE grant DE-SC0013642.
\end{acknowledgements}

\bibliographystyle{utphys}
\bibliography{main}

\end{document}